\documentclass[aps,prc,superscriptaddress,twocolumn]{revtex4}
\usepackage{latexsym}
\usepackage{amsmath}  
\usepackage{amsfonts} 
\usepackage{newtxtext,newtxmath} 
\usepackage{bm} 
\usepackage{graphicx} 
\usepackage{float} %
\usepackage{subfigure}
\usepackage{epsf} %
\usepackage{epstopdf} %
\usepackage{nameref} %
\usepackage[colorlinks,
            citecolor=blue,
            anchorcolor=red,
            menucolor=red,
            linkcolor=red,
            filecolor=red,
            runcolor=red,
            urlcolor=blue,
            frenchlinks=true]{hyperref}
\usepackage{makecell} %
\usepackage{array} %
\usepackage{xcolor} 

\begin{document}

\title{Higher-Order Symmetry Energy Effects on the Properties of Neutron Stars}

\author{Zijing Zhou}
\affiliation{Key Laboratory for Microstructural Material Physics of Hebei Province, School of Science, Yanshan University, Qinhuangdao 066004, China}

\author{Xuhao Wu}\email{wuhaoysu@ysu.edu.cn}
\affiliation{Key Laboratory for Microstructural Material Physics of Hebei Province, School of Science, Yanshan University, Qinhuangdao 066004, China}
\author{Shishao Bao}\email{bao_shishao@163.com}
\affiliation{School of Physics and Electronic Engineering, Shanxi Normal University, Taiyuan 030031, China}
\author{Min Ju}\email{jumin@upc.edu.cn}
\affiliation{School of Science, China University of Petroleum (East China), Qingdao 266580, China}
\author{Jinniu Hu}\email{hujinniu@nankai.edu.cn}
\affiliation{School of Physics, Nankai University, Tianjin 300071, China}
\author{Hong Shen}\email{songtc@nankai.edu.cn}
\affiliation{School of Physics, Nankai University, Tianjin 300071, China}

\begin{abstract}

We study the effects of the symmetry-energy curvature $K_{\rm sym}$ on the equation of state (EoS) and neutron-star properties within the relativistic mean-field (RMF) framework.
We construct controlled families of RMF parametrizations based on the TM1 and IUFSU models by varying $K_{\rm sym}$ while keeping the symmetry energy $J$ and its slope $L$ fixed at saturation density.
We find that, although a larger $K_{\rm sym}$ leads to a lower symmetry energy at suprasaturation densities and a reduced proton fraction in $\beta$-equilibrated matter, 
it results in a stiffer EoS and a higher sound speed.
The maximum neutron-star mass is weakly affected, whereas the canonical radius $R_{1.4}$ exhibits a nearly linear dependence on $K_{\rm sym}$.
 
\end{abstract}

\maketitle  

\section{Introduction}
\label{sec:Intro}

Neutron stars provide a unique laboratory for studying strongly interacting matter under extreme conditions, 
where the baryonic matter is highly isospin asymmetric and can reach densities several times nuclear saturation density~\cite{Lattimer2007,Steiner2005,Lattimer2021}.
Their macroscopic properties are governed by the equation of state (EoS) of dense nuclear matter, 
making neutron-star observations an important probe of the high-density EoS~\cite{Lattimer2001,Lattimer2007}.
In particular, pulsar timing measurements based on the relativistic Shapiro delay have provided precise masses for several massive neutron stars,
including PSR J1614-2230~\cite{Arzoumanian2018}, PSR J0348+0432~\cite{Antoniadis2013}, and PSR J0740+6620~\cite{Riley2021}, 
establishing a robust lower bound approximately 2~$M_\odot$ on the maximum mass supported by the EoS.
Complementary constraints on the mass-radius relation have been obtained from X-ray pulse-profile modeling by the Neutron Star Interior Composition Explorer (NICER), 
including observations of PSR J0030+0451~\cite{Miller2019,Riley2019,Vinciguerra2024}, PSR J0740+6620~\cite{Miller2021,Riley2021,Salmi2024a}, 
PSR J0437-4715~\cite{Choudhury2024}, PSR J1231-1411~\cite{Salmi2024b} and PSR J0614-3329~\cite{Mauviard2025,Miller2026}.
Moreover, the binary neutron-star merger GW170817 has provided constraints on the tidal deformability and, consequently, on the radius of neutron stars~\cite{Abbott2018,Abbott2019,Malik2018}.
Numerous theoretical studies have investigated the connection between the nuclear EoS and neutron-star properties using different nuclear many-body approaches and nucleon-nucleon interactions, 
including effects associated with nuclear symmetry energy~\cite{Margueron2018,Wu2019,Ji2019,Hu2020,Wu2021,Kubis2023,Wu2026}.

The nuclear symmetry energy and its density dependence play an important role in understanding the properties of
neutron-rich matter in both nuclear physics and astrophysics~\cite{Li2008,Tsang2012,Lattimer2014,Baldo2016,Yeasin2025}.
Around the nuclear saturation density, the symmetry energy $E_{\rm sym}$ can be characterized by its value $J$, slope $L$, and curvature $K_{\rm sym}$ at the saturation density $n_0$,
 which describe its density dependence at progressively higher orders.
The symmetry energy at saturation $J$ and its slope $L$ have been extensively investigated using nuclear structure data, 
laboratory experiments, microscopic many-body calculations, and neutron-star observations~\cite{Li2008,Bao2014,Tews2017,Zhang2018,Li2019,Krastev2023,Burgio2024,Ju2025}.
In comparison, the curvature parameter $K_{\rm sym}$ remains substantially less constrained. 
Moreover, different theoretical approaches can yield appreciably different correlations among $J$, $L$, and $K_{\rm sym}$,
leading to significant model dependence in the extrapolation of the symmetry energy to suprasaturation densities~\cite{Holt2018,Li2019,Lattimer2023}.
Since $K_{\rm sym}$ directly characterizes the curvature of $E_{\rm sym}$ around saturation, 
it can substantially influence the high-density behavior of the symmetry energy.
Understanding how $K_{\rm sym}$ affects the high-density symmetry energy and propagates into the equation of state and neutron-star observables is therefore of particular interest.

A variety of complementary approaches have been employed to constrain the density dependence of the nuclear symmetry energy~\cite{Lattimer2023}. 
At sub-saturation and near-saturation densities, information is obtained from nuclear masses, neutron-skin thicknesses, dipole polarizabilities, isovector giant resonances, and heavy-ion collision observables~\cite{Tews2017,An2023,Koliogiannis2026}.
Microscopic calculations of pure neutron matter, particularly those based on chiral effective field theory, 
provide additional theoretical constraints on the density dependence of the symmetry energy~\cite{Fattoyev2012,Kruger2013}.
These terrestrial and microscopic approaches have substantially improved the knowledge of $E_{\rm sym}$ and $L$, 
although their sensitivity to higher-order parameters such as $K_{\rm sym}$ is generally weaker and often correlated with the lower-order symmetry-energy parameters~\cite{Lattimer2023}. 
At suprasaturation densities, neutron-star observations offer an important complementary probe because the radii, tidal deformabilities, and maximum masses of neutron stars depend on the pressure and composition of dense neutron-rich matter~\cite{Lattimer2023}. 
Studies using multimessenger observations, including GW170817 and NICER, have shown that neutron-star observables can provide information on the density dependence of the symmetry energy, 
including $L$ and $K_{\rm sym}$, although the inferred constraints remain dependent on the adopted EoS framework and correlations among model parameters~\cite{Dong2025,Li2020,Li2021,Li2026}.
These considerations highlight the importance of constructing controlled sets of nuclear models in which the effects of individual symmetry-energy parameters can be investigated systematically.

In this work, we investigate the effects of the symmetry-energy curvature $K_{\rm sym}$ on dense nuclear matter and neutron-star properties within the relativistic mean-field (RMF) framework.
To isolate the effects of $K_{\rm sym}$, we construct controlled families of RMF parametrizations 
in which the selected isoscalar saturation properties and the symmetry-energy parameters $J$ and $L$ are kept fixed while $K_{\rm sym}$ is systematically varied.
We adopt the TM1~\cite{sugahara1994relativistic} and IUFSU~\cite{fattoyev2010relativistic} parametrizations as two representative baseline models,
which have different isovector properties and therefore provide complementary frameworks for examining the model dependence of the $K_{\rm sym}$ effects.
For each baseline model, the isovector-scalar $\delta$-meson coupling is introduced,
together with adjustments of the isovector $\rho$-meson couplings to generate the desired values of $K_{\rm sym}$ while preserving the prescribed $J$ and $L$.
This controlled construction allows the effects of varying $K_{\rm sym}$ to be examined with reduced interference from variations in the lower-order symmetry-energy parameters.
We first investigate how $K_{\rm sym}$ modifies the density dependence of the symmetry energy and the proton fraction of $\beta$-equilibrated matter.
We then examine its effects on the pressure, sound speed, and neutron-star mass-radius relation. 
Particular attention is paid to the correlation between the canonical neutron-star radius $R_{1.4}$ and $K_{\rm sym}$,
 and to how this correlation depends on the underlying RMF parametrization when $L$ is fixed. 
 We find that $R_{1.4}$ exhibits an approximately linear dependence on $K_{\rm sym}$, 
 with remarkably similar slopes for the TM1- and IUFSU-based model families for $L=40$~MeV.


This paper is organized as follows. 
In Sec.~\ref{sec:Theory}, we briefly introduce the RMF model and discuss the parameters to be used in this study.
In Sec.~\ref{sec:results}, we show the numerical results and the density dependence of the symmetry energy in neutron-star matter as well as the effects of higher-order symmetry-energy parameter on the neutron-star properties. 
Sec.~\ref{sec:conclusion} is devoted to the conclusions.

\section{formalism}
\label{sec:Theory}
We describe uniform nuclear matter within the RMF framework,
 in which nucleons interact through the exchange of scalar (\(\sigma,\delta\)) and vector (\(\omega,\rho\)) mesons. 
The \(\sigma\) and \(\omega\) fields determine the isoscalar sector, 
whereas the \(\rho\) and \(\delta\) fields provide the isovector degrees of freedom relevant to the density dependence of the symmetry energy.
In this work, the $\delta$ meson is introduced on the basis of the widely validated RMF parametrizations TM1\cite{sugahara1994relativistic} and IUFSU\cite{fattoyev2010relativistic}. 
The Lagrangian density is written as
\begin{eqnarray}
\label{eq:RMF} 
\mathcal{L_{\rm{RMF}}}
& = & \sum_{i=n,p}\bar{\psi}_i \bigg\{i \gamma_{\mu}\partial^{\mu}- (M +g_{\sigma}\sigma) \notag\\
&&  -\gamma_{\mu} \left[g_{\omega}\omega^{\mu} 
   +\frac{g_{\rho}}{2}\tau_a\rho^{a\mu} \right]-g_{\delta}\tau_a\delta^{a} \bigg\}\psi_i  \notag \\
&& +\frac{1}{2}\partial_{\mu}\sigma\partial^{\mu}\sigma -\frac{1}{2}%
m^2_{\sigma}\sigma^2-\frac{1}{3}g_{2}\sigma^{3} -\frac{1}{4}g_{3}\sigma^{4} \notag \\
&& -\frac{1}{4}W_{\mu\nu}W^{\mu\nu} +\frac{1}{2}m^2_{\omega}\omega_{\mu}%
\omega^{\mu}+ \frac{1}{4}c_3(\omega_{\mu}\omega^{\mu})^2 \notag \\
&& -\frac{1}{4}R^a_{\mu\nu}R^{a\mu\nu} +\frac{1}{2}m^2_{\rho}\rho^a_{\mu}%
\rho^{a\mu} \notag \\
&& +\frac{1}{2}\partial_{\mu}\delta^{a}\partial^{\mu}\delta^{a} 
- \frac{1}{2}m^2_{\delta}\delta^{a}\delta^{a} \notag \\
&& + \Lambda_V g_\omega^2 g_\rho^2 (\omega_{\mu}\omega^{\mu})(\rho^a_{\mu}\rho^{a\mu}) \notag  \\
&& +\sum_{l=e,\mu}\bar{\psi}_{l}
\left( i\gamma_{\mu }\partial^{\mu }-m_{l}\right)\psi_l,
\end{eqnarray}  
where $W_{\mu\nu}$ and $R_{\mu\nu}$ denote the antisymmetric tensor fields of the $\omega$ and $\rho$ mesons, respectively.
Within the mean-field approximation, the meson fields are replaced by their classical expectation values,
\begin{equation}
	\begin{aligned}
	\sigma \rightarrow \langle \sigma \rangle &\equiv \sigma, \\
	\omega_{\mu} \rightarrow \langle \omega_{\mu} \rangle &\equiv \omega,\\
	\rho_{\mu} \rightarrow \langle \rho_{\mu} \rangle &\equiv \rho,\\
	\delta \rightarrow \langle \delta \rangle &\equiv \delta.
\end{aligned}
\end{equation}
The resulting Dirac equation for nucleons is
\begin{equation}
	[i\gamma_{\mu}\partial^{\mu}-M^{*}-g_{\omega}\gamma^{\mu}\omega_{\mu}-\frac{g_{\rho}}{2}\tau_a\gamma_{\mu}\rho^{a\mu}]\psi=0,
\end{equation}
where \(M^*\) denotes the isospin-dependent effective mass.
\(\delta\) meson coupling induces an effective-mass splitting between protons and neutrons,
\begin{align}
	M_{p}^{*}=M+g_{\sigma}\sigma+g_{\delta}\delta, \\
	M_{n}^{*}=M+g_{\sigma}\sigma-g_{\delta}\delta.
\end{align}
The meson field equations obtained from the Euler-Lagrange equations reduce to
\begin{equation}
	\begin{aligned}
		m_{\sigma}^{2}\sigma+g_{2}\sigma^{2}+g_{3}\sigma^{3}&=-g_{\sigma}(n_{p}^{s}+n_{n}^{s}), \\
		m_{\omega}^{2}\omega+c_{3}\omega^{3}+2\Lambda_{\mathrm{V}}g_{\omega}^{2}g_{\rho}^{2}\omega\rho^{2}&=g_{\omega}(n_{p}+n_{n}),\\
		m_{\rho}^{2}\rho+2\Lambda_{\mathrm{V}}g_{\omega}^{2}g_{\rho}^{2}\omega\rho^{2}&=\frac{g_{\rho}}{2}(n_{p}-n_{n}),\\
		m_{\delta}^{2}\delta&=g_{\delta}(n_{p}^{s}-n_{n}^{s}),
	\end{aligned}
\end{equation}
where $n_{p}$ and $n_{n}$ are the number densities, and $n_{p}^{s}$ and $n_{n}^{s}$ are the scalar densities, for protons and neutrons, respectively.  
The vector and scalar densities are given by
\begin{align}
	n_{i}^{s}&=\frac{1}{\pi ^{2}}\int_{0}^{k_{F,i}}\frac{M_{i}^{*}}{\sqrt{k^2+M_i^{*2}} }k^2dk, \\
	n_{i}&=\frac{k_{F,i}^3}{3\pi^2}, 
\end{align}
with ${k_{F,i}}$ the Fermi momentum.

The chemical potentials of baryons can be expressed as
\begin{eqnarray}
	\mu_p &=& \sqrt{k_{F,p}^2 + M_p^{*2}} + g_\omega \omega + \frac{g_\rho}{2} \rho, \\
	\mu_n &=& \sqrt{k_{F,n}^2 + M_n^{*2}} + g_\omega \omega - \frac{g_\rho}{2} \rho,
\end{eqnarray}
respectively.
For neutron star matter containing baryons and leptons, the conditions of $\beta$ equilibrium and charge neutrality must also be introduced:
\begin{eqnarray}
\label{eq:beta}
	\mu_n &=&\mu_p+\mu_e, \\
	\mu_e &=&\mu_{\mu}, \\
	n_p &=& n_e+n_{\mu}.
\end{eqnarray}
The conditions of $\beta$ equilibrium and charge neutrality determine the composition of neutron-star matter.

The energy density  $\varepsilon$ and pressure $P$  are then obtained as
\begin{eqnarray}
	\varepsilon &=& \sum_{i=n,p} \frac{1}{\pi^2} \int_0^{k_{F,i}} \sqrt{k^2 + M_i^{*2} k^2} dk \notag\\
	&& + \frac{1}{2} m_\sigma^2 \sigma^2 + \frac{1}{3} g_2 \sigma^3 + \frac{1}{4} g_3 \sigma^4 
	+ \frac{1}{2} m_\omega^2 \omega^2+\frac{3}{4}c_3\omega^4    \notag\\
	&&+ \frac{1}{2} m_\rho^2 \rho^2+\frac{1}{2} m_\delta^2 \delta^2+3\Lambda_{\mathrm{V}}g_{\omega}^2g_{\rho}^2\omega^2\rho^2 \notag\\
	&& +\sum_{l=e,\mu}\frac{1}{\pi^2}\int_{0}^{k_{F,l}} k^2\sqrt{k^2+m_l^2}dk,
\end{eqnarray}

\begin{eqnarray}
	P &=& \sum_{i=n,p} \frac{1}{3\pi^2} \int_0^{k_{F,i}} \frac{1}{\sqrt{k^2 + M_i^{*2}}} k^4 dk \notag\\
	&& - \frac{1}{2} m_\sigma^2 \sigma^2 - \frac{1}{3} g_2 \sigma^3 - \frac{1}{4} g_3 \sigma^4 
	+ \frac{1}{2} m_\omega^2 \omega^2+\frac{1}{4}c_3\omega^4  \notag\\
	&& + \frac{1}{2} m_\rho^2 \rho^2-\frac{1}{2} m_\delta^2 \delta^2+\Lambda_{\mathrm{V}}g_{\omega}^2g_{\rho}^2\omega^2\rho^2  \notag\\
	&&   + \sum_{l=e,\mu} \frac{1}{3\pi^2} \int_0^{k_{F,l}} \frac{1}{\sqrt{k^2 + m_l^{2}}} k^4 dk.
\end{eqnarray}

Near the saturation density, the symmetry energy can be expanded in terms of density:
\begin{align}
	E_{\mathrm{sym}}=J+Lx+\frac{1}{2}K_{\mathrm{sym}}x^2+\mathcal{O}(x^3),
	\label{eq:esym}
\end{align}
here, $x=(n_b-n_0)/(3n_0)$ is the dimensionless density variable commonly used in the expansion around saturation density $n_0$,
the symmetry energy $J$, the symmetry energy slope $L$ and the symmetry energy curvature $K_{\mathrm{sym}}$ at saturation density are given respectively by
\begin{align}
	\label{eq:j}
	J &\equiv E_{\mathrm{sym}}({n_0}), \\
	\label{eq:l}
	L&=3n_0\frac{\partial E_{\mathrm{sym}}}{\partial n_b}\Bigg|_{n_b=n_0}, \\
	K_{\mathrm{sym}}&=9n_0^2\frac{\partial^2 E_{\mathrm{sym}}}{\partial n_b^2}\Bigg|_{n_b=n_0}.
	\label{eq:k}
\end{align}

In order to examine the influence of the high-order symmetry-energy curvature \(K_{\rm sym}\), 
we generate two sets of models based on the TM1~\cite{sugahara1994relativistic} and
IUFSU~\cite{fattoyev2010relativistic} parametrizations.
The two parametrizations differ in their isoscalar and isovector sectors and therefore provide complementary baselines for examining the model dependence of the \(K_{\rm sym}\) effects.
The parameters are listed in Table \ref{tab:IUFSU_TM1}.
The TM1 parametrization has $J$=36.9 MeV, with $L=110.8$ MeV and $K_{\mathrm{sym}}$=33.6 MeV, 
while the IUFSU model gives $J$=31.3 MeV, $L$=47.2 MeV, and $K_{\mathrm{sym}}$=28.6 MeV. 
For each baseline parametrization, the isoscalar-sector parameters are kept unchanged, 
while the three isovector couplings $g_\rho$, $g_\delta$, and $\Lambda_V$ are adjusted simultaneously.
For a prescribed set of symmetry-energy parameters $(J,L,K_{\rm sym})$,
 these couplings are determined by imposing the corresponding values of the symmetry energy, its slope, and its curvature at saturation density. 
 In this way, $K_{\rm sym}$ can be varied while $J$ and $L$ are held fixed, allowing its effects on the high-density equation of state and neutron-star observables to be isolated.
For the TM1 model, we constrain $K_{\mathrm{sym}}$ within the range of approximately $-120$~MeV to $-70$~MeV while maintaining $L=40$~MeV. 
For the IUFSU model, $K_{\mathrm{sym}}$ is constrained within the range of approximately -10~MeV to 80 MeV while maintaining $L=40$~MeV, same as the TM1 family.
These ranges of $K_{\mathrm{sym}}$ are consistent with those reported in Refs~\cite{Mondal2017,Li2020,Li2021,Dong2025},
though most studies prefer negative $K_{\mathrm{sym}}$~\cite{Li2021}.

The coupling constants $g_{\rho}$, $g_{\delta}$, $\Lambda_{\mathrm{V}}$ and the corresponding $K_{\mathrm{sym}}$ values for each parametrization built on the TM1 and IUFSU models are listed in Table \ref{tab:TM1-Ksym} and \ref{tab:IUFSU-Ksym}, respectively.
\begin{table*}
	\centering  %
	\hspace{0pt}
	\setlength{\tabcolsep}{1.5mm}   %
		\addtolength{\extrarowheight}{5pt} %
	\caption{Parameters of the baseline TM1 and IUFSU parametrizations~\cite{sugahara1994relativistic,fattoyev2010relativistic}. 
	The masses are given in MeV.}
	\begin{tabular}{cccccccccccccc}
		\hline\hline
		Model & $M$ & $m_\sigma$ & $m_\omega$ & $m_\rho$ & $m_\delta$ & $g_\sigma$ & $g_\omega$ & $g_\rho$ & $g_\delta$ & $g_2$ (fm$^{-1}$) & $g_3$ & $c_3$ & $\Lambda_{\mathrm{V}}$  \\
		\hline
		IUFSU & 939.000 & 491.500 & 782.500 & 763.000 & 980.000 & 9.971 & 13.032 & 13.590 & 0.000 & $-$8.493 & 0.488 & 144.220 & 0.046 \\
		TM1 & 938.000 & 511.198 & 783.000 & 770.000 & 980.000 & 10.029 & 12.614 & 9.264 & 0.000 & $-$7.233 & 0.618 & 71.308 & 0.000 \\
		\hline\hline
	\end{tabular}
	\label{tab:IUFSU_TM1}
\end{table*}

\begin{table}
	\centering  
	\setlength{\tabcolsep}{0.7mm}   %
	\addtolength{\extrarowheight}{5pt} %
	\caption{Values of the isovector coupling constants $g_\rho$, $g_\delta$, and $\Lambda_{\mathrm{V}}$ generated from the TM1 parametrization for different values of the symmetry-energy curvature $K_{\rm sym}$,
	with the symmetry-energy slope fixed at $L=40$~MeV and the symmetry energy at saturation density fixed at $J=36.91$~MeV.}
	\begin{tabular}{ccccccc}
		\hline\hline
		$K_{\mathrm{sym}}$(MeV) & $-$120 & $-$110 & $-$100 & $-$90 & $-$80 & $-$70 \\
		\hline
		$g_\rho$ & 15.6887 & 15.5300 & 15.4211 & 15.3526 &  15.3187 & 15.3162  \\
		$g_\delta$ & 4.5225 &  4.0905 & 3.6866 & 3.2976 &   2.9128 & 2.5208 \\
		$\Lambda_{\mathrm{V}}$ & 0.0147 & 0.0163 & 0.0179 & 0.0196 &  0.0212 & 0.0229  \\
		\hline\hline
	\end{tabular}
	\label{tab:TM1-Ksym}
\end{table}

\begin{table}
	\centering  
	\setlength{\tabcolsep}{0.7mm}   %
	\addtolength{\extrarowheight}{5pt} %
		\caption{Values of the isovector coupling constants $g_\rho$, $g_\delta$, and $\Lambda_{\mathrm{V}}$ generated from the IUFSU parametrization for different values of the symmetry-energy curvature $K_{\rm sym}$,
	with the symmetry-energy slope fixed at $L=40$~MeV and the symmetry energy at saturation density fixed at  $J$ = 31.28~MeV.}
	\begin{tabular}{ccccccc}
		\hline\hline
		$K_{\mathrm{sym}}$(MeV) & $-$10 & 0 & 20 & 40 & 60 & 80 \\
		\hline
		$g_\rho$ & 14.2698 & 14.1604 & 14.2696 & 14.7029 & 15.4828 &16.7480  \\
		$g_\delta$ & 4.5492 & 4.0201 & 3.1968 & $-$2.4851 &  1.7675 & 8.3293  \\
		$\Lambda_{\mathrm{V}}$ & 0.0193 & 0.0229 & 0.0296 & 0.0364 & 0.0433 & 0.0506  \\
		\hline\hline
	\end{tabular}
	\label{tab:IUFSU-Ksym}
\end{table}

\section{Results and discussion}
\label{sec:results}

\begin{figure}[tbhp]
	\hspace{0pt}
	\includegraphics[width=1.0\linewidth]{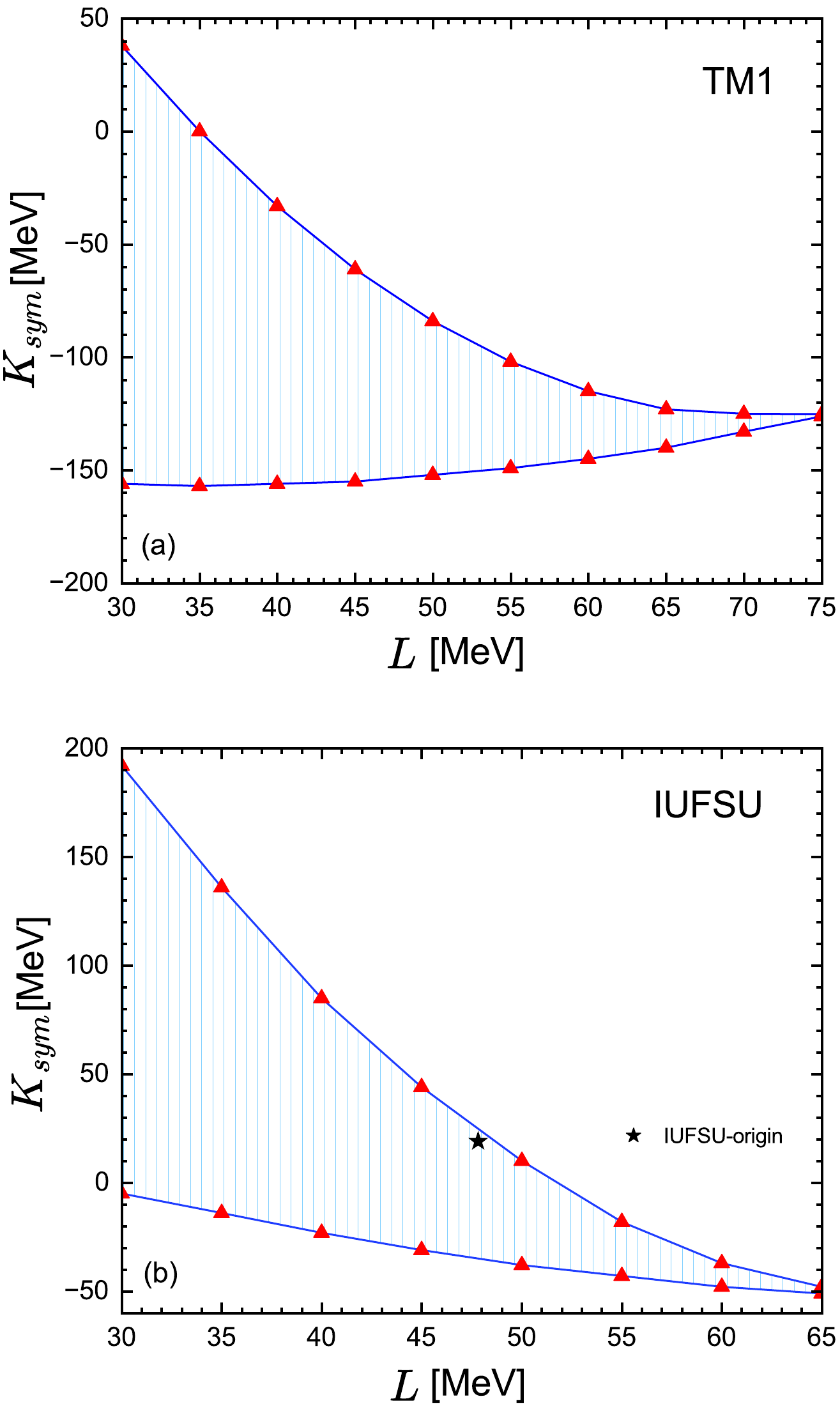}%
	\caption{ Symmetry energy curvature parameter $K_{\mathrm{sym}}$ as a function of the symmetry energy slope $L$ for the TM1 (upper panel) and IUFSU (lower panel) parametrizations. As $L$ increases, the allowed range of $K_{\mathrm{sym}}$ shrinks and reaches its limit at $L \approx 75$ (65) MeV for TM1 (IUFSU). }
	\label{fig:1-L-K}
\end{figure}

In this section, we investigate the influence of the symmetry energy curvature $K_{\mathrm{sym}}$ on the neutron star EoS and macroscopic properties. 
Firstly, we construct two sets of models within the RMF framework, based on the TM1 and IUFSU parametrizations.
For each choice of the symmetry-energy parameters $(J, L, K_{\mathrm{sym}})$, 
the three isovector coupling constants $g_{\rho}$,  $\Lambda_{\mathrm{V}}$ and $g_{\delta}$ are determined by simultaneously imposing the saturation values of the symmetry energy, its slope, and its curvature, i.e., by solving Eqs.~(\ref{eq:j}-\ref{eq:k}).
Because these conditions constitute a system of nonlinear coupled equations, 
real and non-zero solutions exist only within a finite region of the $(L, K_{\mathrm{sym}})$ plane with a given $J$.
Fig.~\ref{fig:1-L-K} shows the boundaries of this allowed $(L, K_{\mathrm{sym}})$ region for the TM1 (upper panel) and IUFSU (lower panel) parametrizations.
the original IUFSU parametrization is labeled by a black star, the original TM1 point ($L=110.8$~MeV) lies outside the plotted range.
Overall, as $L$ increases, the allowed interval of $K_{\mathrm{sym}}$ narrows monotonically.
The upper boundary of $K_{\mathrm{sym}}$  decreases rapidly with increasing $L$,
 whereas the lower boundary varies only mildly and shows opposite trends in the two models,
 which rises slightly with $L$  for TM1, while decreases slightly for IUFSU. 
 The asymmetric evolution of the two boundaries leads to a continuous shrinking of the solution domain.
 Beyond a critical slope, $L\sim75$~MeV for TM1 and $L\sim65$~MeV for IUFSU, the allowed region collapses entirely,
 meaning that no combination of $g_{\rho}$,  $\Lambda_{\mathrm{V}}$ and non-zero $g_{\delta}$ can simultaneously reproduce the prescribed $J$, $L$, and $K_{\mathrm{sym}}$ while preserving the isoscalar saturation properties.
 With $g_{\delta}=0$, there still exist real solutions with larger $L$.
This behavior reflects the fact that the adjustment space of the isovector sector is progressively exhausted as $L$ grows,
the slope $L$ is controlled mainly by $g_{\rho}$ and $\Lambda_{\mathrm{V}}$, 
and once $L$ is large, the remaining freedom in $g_{\delta}$ is insufficient to accommodate an arbitrary $K_{\mathrm{sym}}$.
The $L$-$K_{\rm sym}$ range found in our model construction is qualitatively consistent with the correlations reported in previous studies using RMF and other models~\cite{Mondal2017,Li2020,Li2021,Dong2025}.
However, the detailed form and strength of the correlation vary considerably among different model families, and systematic comparisons have demonstrated a substantial model dependence of $K_{\rm sym}$ at a given $L$~\cite{Li2020}.
Our results further show that, within the present controlled RMF construction, 
this model dependence manifests itself not only in the functional relationship of the $L$-$K_{\rm sym}$ correlation, but also in the finite allowed region and its upper and lower boundaries.

\begin{figure*}[tbhp]
	\hspace{0pt}
	\includegraphics[width=1.0\linewidth]{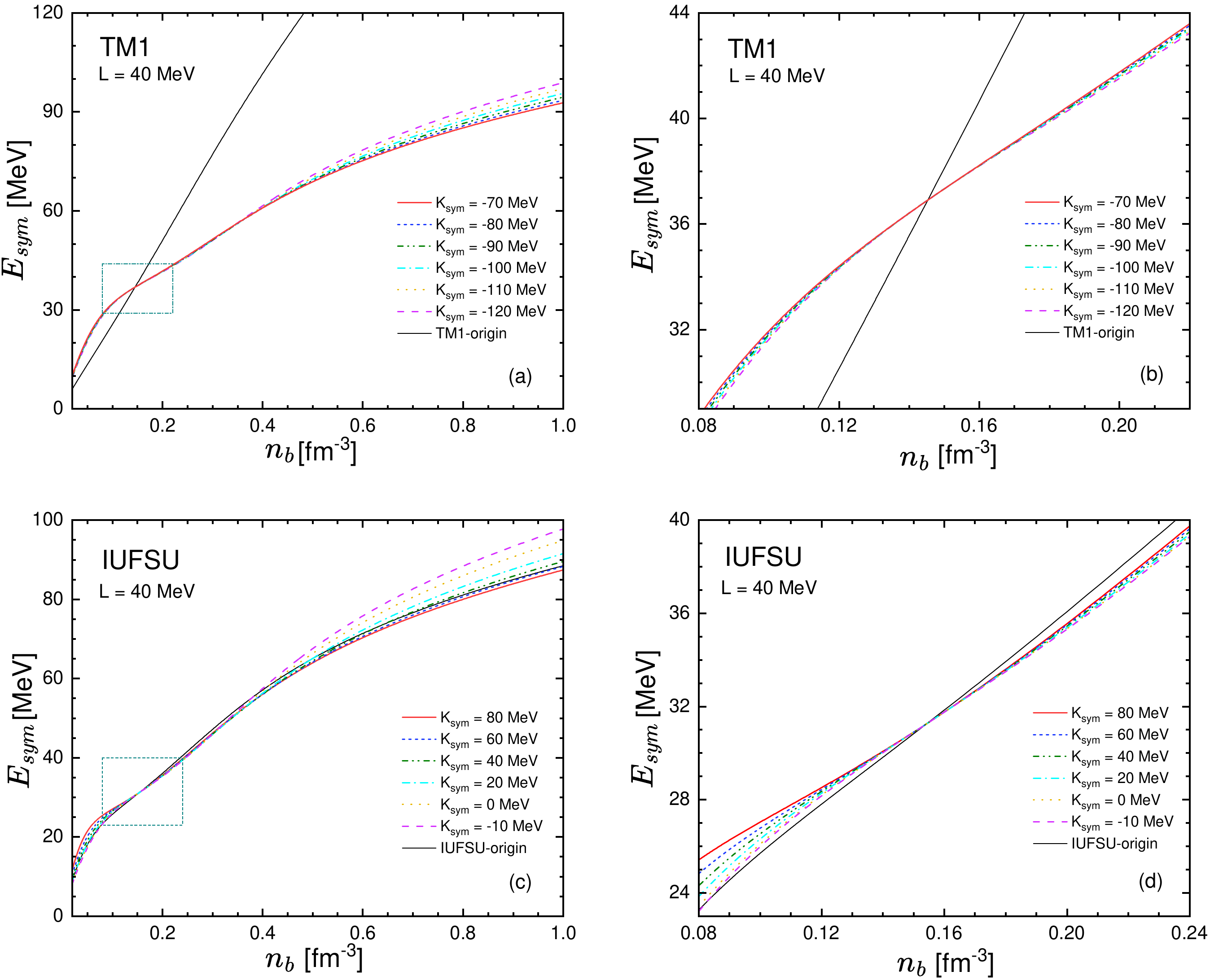}%
	\caption{ Symmetry energy $E_{\mathrm{sym}}$ as a function of baryon density $n_b$ for the TM1 (upper panel) and IUFSU (lower panel) model families with different values of $K_{\mathrm{sym}}$. 
	The right panels show enlarged views of the regions enclosed by the boxes in the left panels.
	 }
	\label{fig:2-nb-esym}
\end{figure*}
\begin{figure*}[tbhp]
	\hspace{0pt}
	\includegraphics[width=1.0\linewidth]{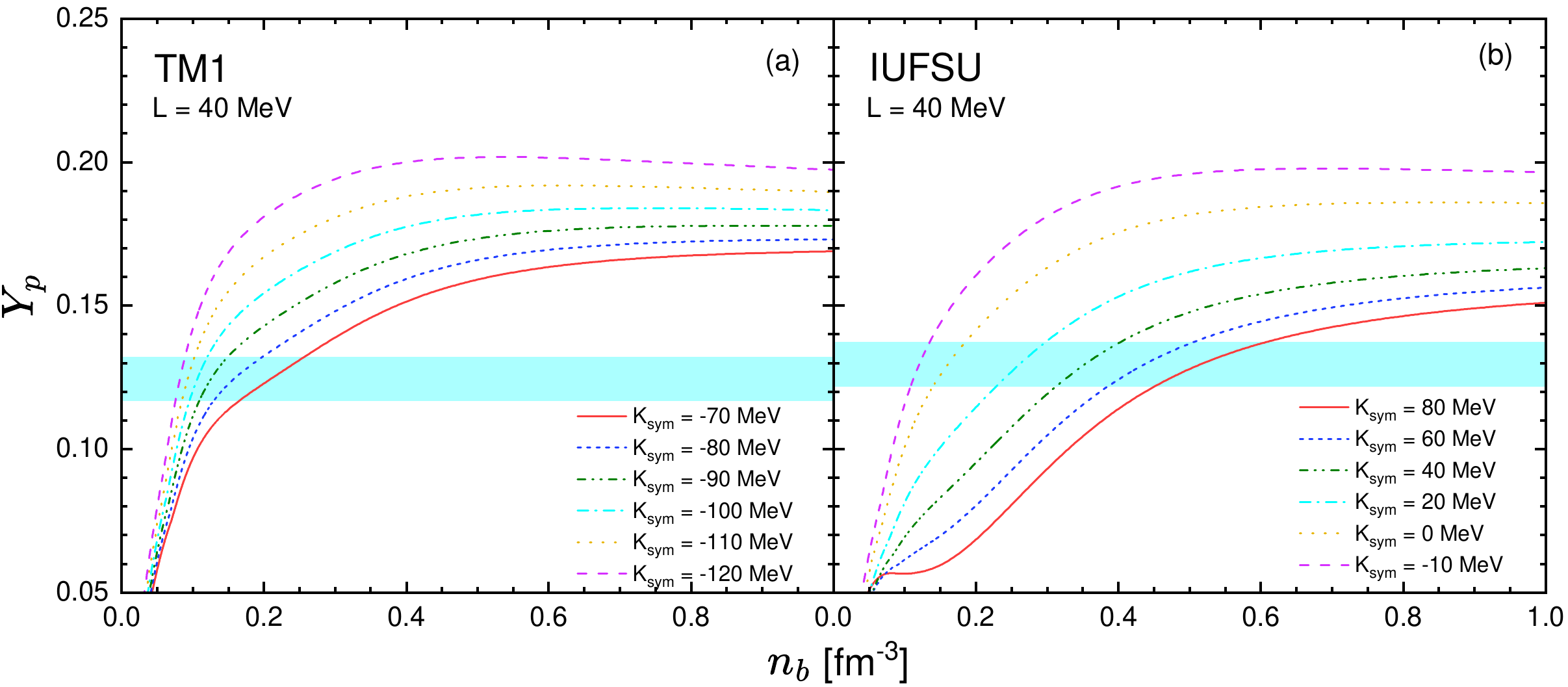}%
	\caption{Proton fraction $Y_p$ as a function of baryon number density $n_b$ for the TM1 (left panel) and IUFSU (right panel) model families with different values of $K_{\mathrm{sym}}$. 
	The cyan bands indicate the ranges of threshold densities for the onset of the direct Urca (DU) process obtained for the different $K_{\mathrm{sym}}$ values. }
	\label{fig:3-nb-yp}
\end{figure*}

Fig.~\ref{fig:2-nb-esym} shows the density dependence of the symmetry energy $E_{\mathrm{sym}}$ for the TM1 (upper panel) and IUFSU (lower panel)  model families constructed with fixed $J$ and $L=40$~MeV but different $K_{\mathrm{sym}}$.
The right panels show enlarged views of the regions around $n_0$.
For both model families, the curves remain very close around saturation density, 
while the dependence on $K_{\rm sym}$ becomes increasingly pronounced at suprasaturation densities.
Small differences among the IUFSU curves are already visible below $n_b\sim0.1$~fm$^{-3}$,
whereas the TM1 curves remain nearly indistinguishable up to $n_b\sim0.3$~fm$^{-3}$.
Above $\sim$0.4~fm$^{-3}$, the spread between different $K_{\mathrm{sym}}$  values grows with density.
In this suprasaturation-density region, a larger algebraic value of $K_{\mathrm{sym}}$ is associated with a smaller $E_{\rm sym}$ for both model families.
At first sight, this behavior appears inconsistent with the local expansion in Eq.~(\ref{eq:esym}), 
which implies that, at fixed $J$ and $L$, a larger $K_{\rm sym}$ increases $E_{\rm sym}$.
The apparent discrepancy arises because Eq.~(\ref{eq:esym}) is a local expansion around $n_0$, 
rather than a complete representation of the symmetry energy at arbitrary density: 
for two models sharing the same $J$ and $L$,
one has $\Delta E_{\mathrm{sym}}(n_b)=\frac{1}{2}\Delta K_{\mathrm{sym}}x^2+\mathcal{O}(x^3)$.
Thus, sufficiently close to $n_0$, a larger $K_{\rm sym}$ necessarily gives a slightly larger $E_{\rm sym}$, as also seen in the enlarged panels.
Indeed, the IUFSU family already exhibits this ordering at very low density.
The reversal of the ordering above $n_b\sim0.4$~fm$^{-3}$ therefore indicates that the net contribution from higher-order terms becomes sufficiently important to overcome the positive second-order contribution associated with a larger $K_{\mathrm{sym}}$.
This behavior highlights the increasing importance of higher-order density dependence when the symmetry energy is extrapolated to suprasaturation densities.
The symmetry-energy curvature $K_{\rm sym}$ affects the density dependence of the symmetry energy $E_{\rm sym}(n_b)$, which in turn influences the composition and thermodynamic properties of $\beta$-equilibrated nuclear matter.

The resulting proton fraction $Y_p$ in $\beta$-equilibrated matter is shown as a function of baryon number density $n_b$ in Fig.~\ref{fig:3-nb-yp}.
The cyan bands indicate the range of densities at which the direct Urca (DU) process is allowed for the different $K_{\rm sym}$ values.
For both model families, the dependence of $Y_p$ on $K_{\rm sym}$ is relatively weak at low densities but becomes increasingly pronounced toward higher densities
 and a larger $K_{\rm sym}$ corresponds to a smaller proton fraction.
Within the parabolic approximation for the isospin dependence of the energy per particle, 
\begin{equation}
\frac{E}{A}(n_b,\alpha) = \frac{E_0}{A}(n_b) + E_{\mathrm{sym}}(n_b)\,\alpha^2,
\end{equation}
 where $\alpha=1-2Y_p$,
the $\beta$-equilibrium Eq.~(\ref{eq:beta}) gives
$\mu_n-\mu_p \simeq 4E_{\rm sym}(n_b)(1-2Y_p)$.
At fixed $n_b$, a smaller $E_{\rm sym}$ therefore corresponds to a lower $Y_p$.
Since Fig.~\ref{fig:2-nb-esym} has shown that a larger $K_{\rm sym}$ leads to a lower $E_{\rm sym}$ in the high-density region, the decrease in $Y_p$ with increasing $K_{\rm sym}$ follows naturally. 
Consequently, the decrease in $Y_p$ with increasing $K_{\rm sym}$ shifts the onset of the DU process to higher densities, as indicated by the cyan bands in Fig.~\ref{fig:3-nb-yp}.

\begin{figure}[tbhp]
	\hspace{0pt}
	\includegraphics[width=1.0\linewidth]{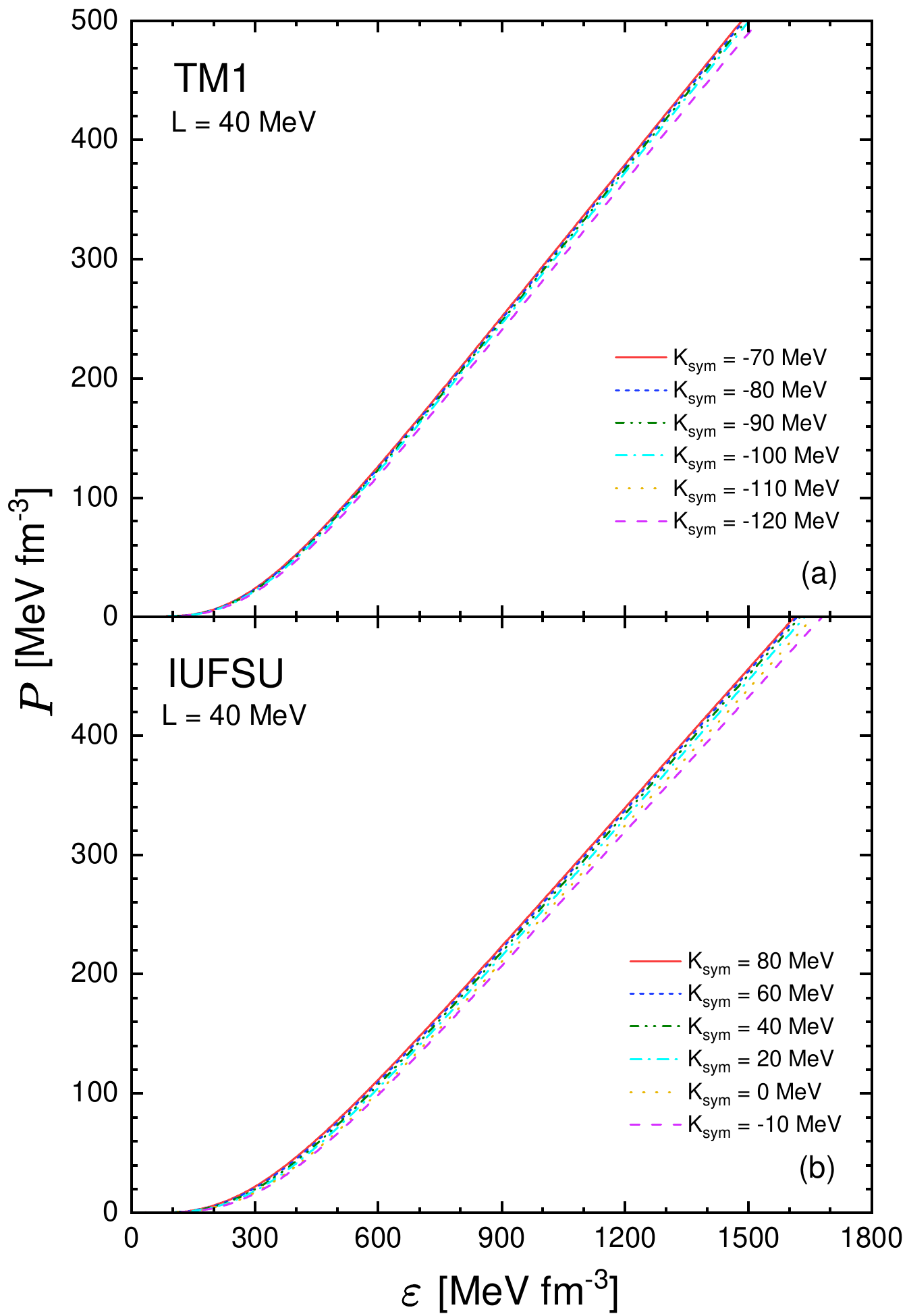}%
	\caption{ Pressure $P$ as a function of energy density $\varepsilon$ for the TM1 (upper panel) and IUFSU (lower panel) model families with different values of the symmetry-energy curvature $K_{\mathrm{sym}}$. }
	\label{fig:4-E-P}
\end{figure}
\begin{figure*}[tbhp]
	\hspace{0pt}
	\includegraphics[width=1.0\linewidth]{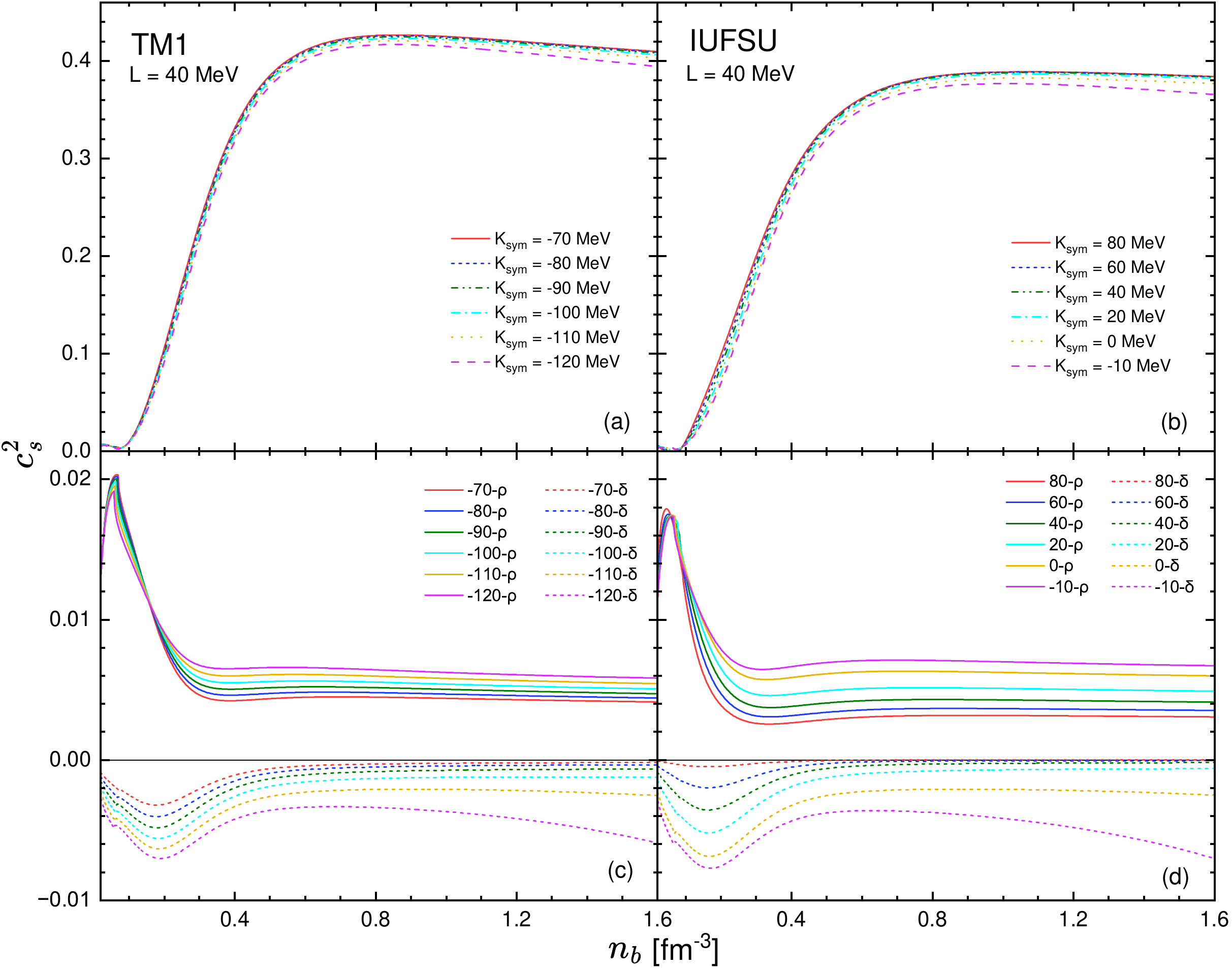}%
	\caption{ Squared sound speed $c_s^2=\partial P/\partial\varepsilon$ as a function of baryon number density $n_b$ for different values of $K_{\rm sym}$ in the TM1 (left panels) and IUFSU (right panels) model families.
	The lower panels show the separate contributions of the $\rho$- and $\delta$-meson channels to the squared sound speed.
	 }
	\label{fig:5-Pnb-cs2}
\end{figure*}

The corresponding pressure-energy density relations for $\beta$-equilibrated matter are shown in Fig.~\ref{fig:4-E-P}.
In both model families, the EoS curves become increasingly separated toward higher energy densities.
At a given energy density, a larger $K_{\rm sym}$ corresponds to a higher pressure, indicating a stiffer EoS, whereas a smaller $K_{\rm sym}$ leads to a softer EoS.
This behavior is closely related to the change in the proton fraction discussed above: 
a larger $K_{\rm sym}$ leads to a smaller $Y_p$ and hence a larger isospin asymmetry in $\beta$-equilibrated matter. 
The resulting increase in the asymmetry of matter enhances the pressure, giving rise to a stiffer EoS for larger $K_{\rm sym}$.
The systematic dependence of the EoS on $K_{\rm sym}$ reflects the modification of the density-dependent interaction in the isovector sector induced by the simultaneous adjustment of \(g_\rho\), \(g_\delta\), and $\Lambda_V$. 
The detailed contributions of the individual isovector channels to the high-density EoS stiffness are discussed below.

Fig.~\ref{fig:5-Pnb-cs2} shows the squared sound speed, $c_s^2=\partial P/\partial\varepsilon$, as a function of baryon number density $n_b$ for different $K_{\rm sym}$. 
The lower panels further display the separate contributions from the $\rho$- and $\delta$-meson channels.
The overall behavior is consistent with the EoS trends shown in Fig.~\ref{fig:4-E-P},
and model parametrizations with larger $K_{\rm sym}$ generally exhibit a larger $c_s^2$ in the high-density region, 
whereas smaller $K_{\rm sym}$ leads to a lower sound speed.
To examine the respective roles of the isovector $\rho$- and $\delta$-meson couplings in the density dependence of the sound speed, we decompose the total pressure into the corresponding mesonic contributions and the free Fermi gas contribution. 
We define the associated contributions to the squared sound speed as
\begin{equation}
\nonumber
c_{s,\rho}^{2}\equiv\frac{dP_{\rho}}{d\varepsilon},
\qquad
c_{s,\delta}^{2}\equiv\frac{dP_{\delta}}{d\varepsilon},
\end{equation}
where $P_{\rho}$ and $P_{\delta}$ denote the pressure contributions associated with the $\rho$- and $\delta$-meson coupling terms, respectively, and $\varepsilon$ is the total energy density of the system. 
These quantities represent a decomposition of the pressure response with respect to the total energy density, rather than independent sound speeds of separate components. 
All derivatives are evaluated along the same $\beta$-equilibrated EoS.
The channel decomposition shows that the $\rho$-meson channel provides a positive contribution to $c_s^2$, 
while the $\delta$-meson channel gives a negative contribution.
The magnitudes of these contributions vary systematically with $K_{\rm sym}$, 
reflecting the corresponding changes in the isovector couplings used to construct the model families.
Toward high densities, both contributions tend to approach saturation for most parametrizations.
An exception occurs for the smallest $K_{\rm sym}$ considered, namely $K_{\rm sym}=-120$~MeV for TM1 and $K_{\rm sym}=-10$~MeV for IUFSU, 
for which the negative $\delta$-meson contribution continues to increase in magnitude at high density.
As a consequence, $c_s^2$ bends downward in the high-density region for these extreme low-$K_{\rm sym}$ parametrizations.
Around the saturation-density region, these two isovector contributions to $c_s^2$ are relatively pronounced where the symmetry energy exhibits a comparatively rapid density evolution.
Since $c_s^2$ involves the derivative of the pressure with respect to the energy density, 
this enhanced density dependence of the symmetry energy leads to a stronger manifestation of the isovector contributions in the sound speed.

\begin{figure*}[tbhp]
	\hspace{0pt}
	\includegraphics[width=1.0\linewidth]{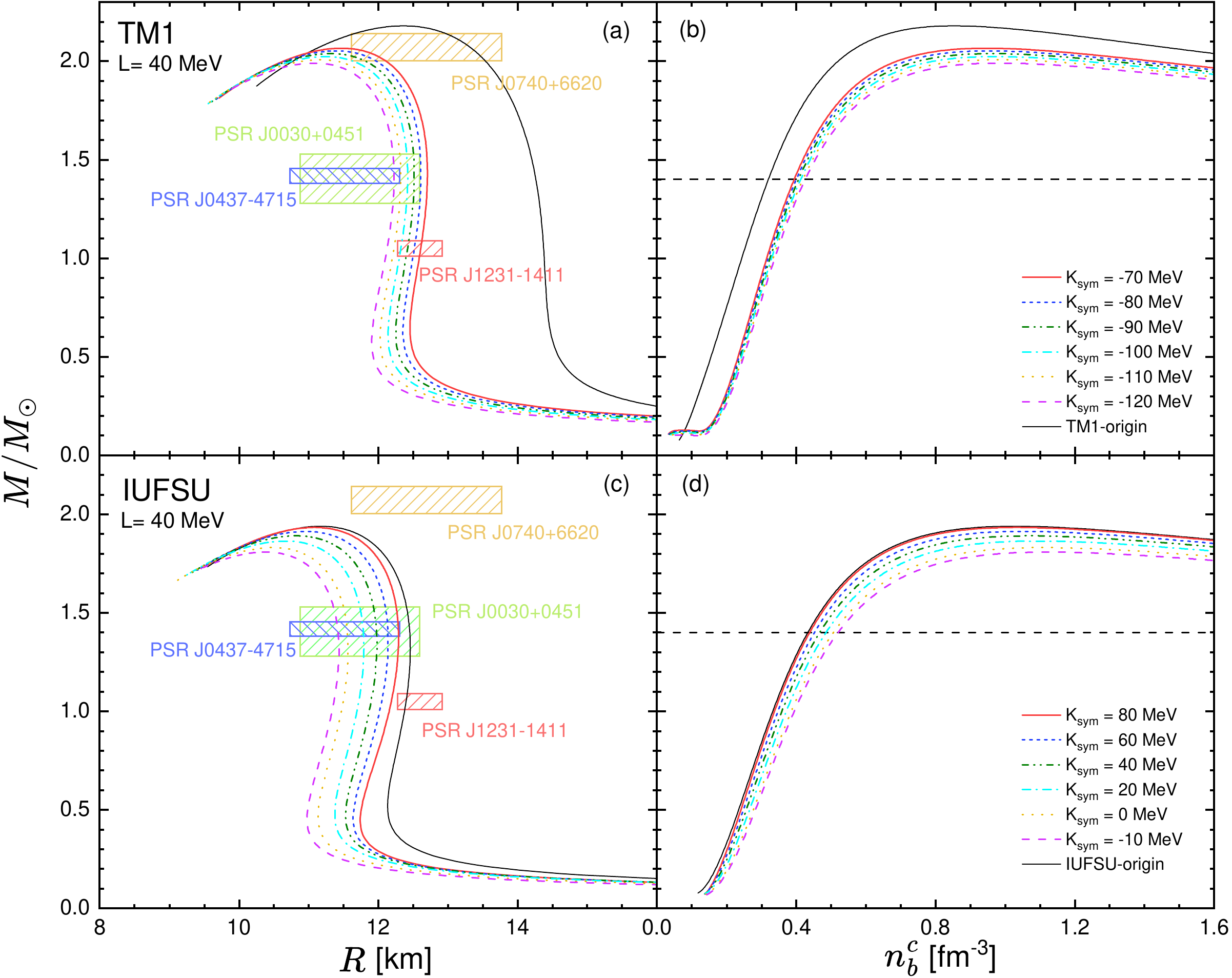}%
	\caption{ Neutron star mass $M$ as a function of radius $R$ (left panels) and central baryon number density $n_b^c$ (right panels) for the TM1 (upper panels) and IUFSU (lower panels) model families with different values of  $K_{\mathrm{sym}}$. The black dashed lines in the right panels indicate the canonical mass $M=1.4\,M_\odot$.}
	\label{fig:6-Rnb-M}
\end{figure*}
\begin{figure}[tbp]
	\hspace{0pt}
	\includegraphics[width=1.0\linewidth]{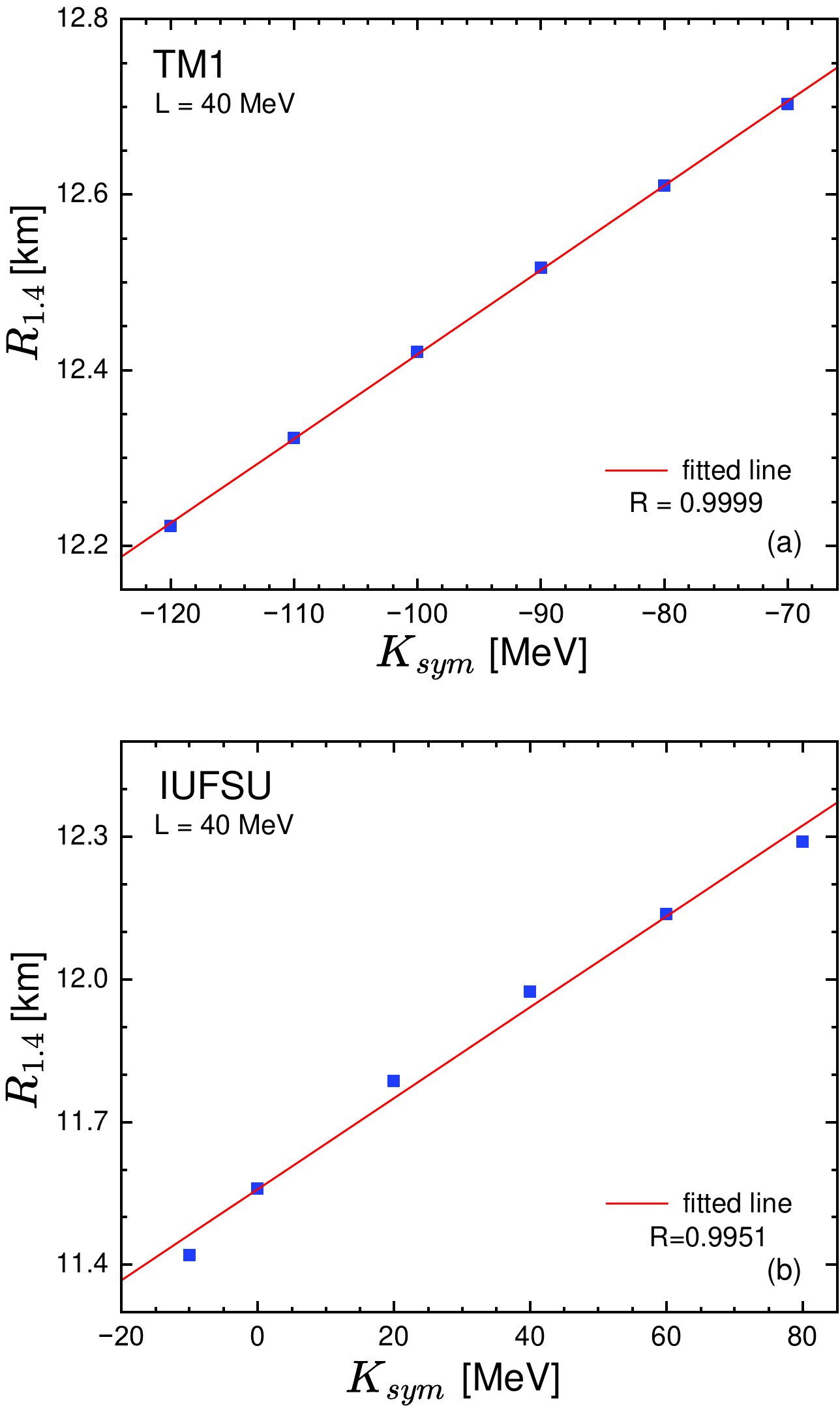}%
	\caption{ Radius $R_{1.4}$ of a $1.4\,M_\odot$ neutron star as a function of the symmetry energy curvature $K_{\mathrm{sym}}$ for the TM1 (upper panels) and IUFSU (lower panels) model families with $L=40$~MeV. The solid lines show the corresponding linear fits. }
	\label{fig:8-Ksym-R1.4}
\end{figure}

Fig.~\ref{fig:6-Rnb-M}  shows the mass-radius relations and the corresponding central baryon number densities of neutron stars obtained from the TM1 and IUFSU model families for different values of $K_{\rm sym}$. 
The calculated mass-radius curves are also compared with representative observational constraints.
The variation of $K_{\rm sym}$ produces only a modest change in the maximum mass, 
whereas the corresponding stellar radii exhibit a more pronounced dependence.
For the TM1 family, $M_{\max}$ varies within $1.99\!-\!2.06\,M_\odot$, while for the IUFSU family it varies within $1.89\!-\!1.93\,M_\odot$. 
In both model families, decreasing $K_{\rm sym}$ leads to smaller radii, corresponding to more compact configurations.
This behavior reflects the cumulative effect of the EoS differences on the stellar structure, 
resulting in more pronounced variations in the mass-radius relation than in the EoS itself, 
with the stellar radius being more sensitive than the maximum mass.
Although $K_{\rm sym}$ modifies the high-density pressure and hence the stiffness of the EoS, 
the resulting change is not sufficient to produce a large variation in the maximum supported mass within the parameter ranges considered here. 
In contrast, the radius of a canonical-mass neutron star provides a more sensitive probe of the changes in the density dependence of the symmetry energy.
The decrease in radius with decreasing $K_{\rm sym}$ is consistent with the corresponding reduction in the high-density pressure and sound speed shown in Figs.~\ref{fig:4-E-P} and~\ref{fig:5-Pnb-cs2}.

To quantify the sensitivity of the canonical neutron-star radius $R_{1.4}$ to the symmetry-energy curvature,
we show $R_{1.4}$ as a function of $K_{\rm sym}$ in Fig.~\ref{fig:8-Ksym-R1.4} for the TM1 and IUFSU model families with $L=40$~MeV. 
A nearly linear correlation is obtained for both model families. 
The corresponding linear fits are
\begin{equation}
\nonumber
R_{1.4}=13.3789+0.00961K_{\rm sym},
\qquad R=0.9999,
\end{equation}
for TM1, and
\begin{equation}
\nonumber
R_{1.4}=11.5589+0.00957K_{\rm sym},
\qquad R=0.9951,
\end{equation}
for IUFSU.
Within the parameter ranges considered here, $R_{1.4}$ increases approximately linearly with $K_{\rm sym}$.
This behavior is qualitatively similar to the well-known correlation between neutron-star radius and the symmetry-energy slope $L$~\cite{Hu2020}.
The nearly linear dependence indicates that, once $J$ and $L$ are fixed, the variation of the high-density EoS induced by $K_{\rm sym}$ can be systematically reflected in the radius of a canonical-mass neutron star.
The linear fits yield nearly identical slopes for the two model families, indicating that the response of $R_{1.4}$ to $K_{\rm sym}$ is remarkably similar between the TM1 and IUFSU constructions with the same $L=40$~MeV.

\section{Conclusions}
\label{sec:conclusion}

In this work, we have investigated the impact of the symmetry-energy curvature $K_{\rm sym}$ on the properties of dense matter and neutron stars within RMF models. 
Starting from the TM1 and IUFSU parametrizations, we constructed model families with different $K_{\rm sym}$ values while keeping the symmetry-energy slope $L=40$ MeV fixed. 
The corresponding isovector couplings, including $g_\rho$, $g_\delta$, and $\Lambda_{\mathrm V}$, were adjusted consistently to generate the desired symmetry-energy parameters.

We find that the effect of $K_{\rm sym}$ becomes increasingly important at suprasaturation densities. 
Although the symmetry energy is nearly insensitive (positively correlated) to $K_{\rm sym}$ around saturation density, 
its high-density behavior shows a pronounced dependence on $K_{\rm sym}$. 
In particular, within the model families considered here, a larger $K_{\rm sym}$ is associated with a smaller symmetry energy at suprasaturation densities. 
This behavior cannot be understood from the quadratic term of the local expansion around saturation alone, indicating the increasing importance of higher-order density-dependent terms at high densities.
A larger $K_{\rm sym}$ leads to a smaller proton fraction at suprasaturation densities.
The EoS becomes stiffer for larger $K_{\rm sym}$, resulting in a higher pressure at a given energy density. 
The analysis of the squared sound speed further shows that the $\rho$-meson and $\delta$-meson channels provide positive and negative contributions, respectively. 
For most parametrizations, these contributions tend to level off at high density, whereas for the extreme low-$K_{\rm sym}$ cases the increasingly negative $\delta$-meson contribution can lead to a downward bending of $c_s^2$ at high density.

For neutron-star properties, the variation of $K_{\rm sym}$ produces only a modest change in the maximum mass, while the stellar radius is more sensitive to the resulting changes in the high-density EoS.
In particular, the canonical radius $R_{1.4}$ exhibits an approximately linear dependence on $K_{\rm sym}$ in both the TM1 and IUFSU model families. 
For $L=40$ MeV, the fitted slopes of $R_{1.4}$ with respect to $K_{\rm sym}$ are remarkably similar in the two RMF model families considered here.

Overall, our results demonstrate that $K_{\rm sym}$, although less constrained than $J$ and $L$, can have a significant impact on the suprasaturation behavior of dense matter and on observable neutron-star properties.
In particular, the nearly model-independent linear response of $R_{1.4}$ to $K_{\rm sym}$ at fixed $L$ suggests that neutron-star radius measurements may provide a useful constraint on the curvature of the symmetry energy and, more generally, on the density dependence of nuclear symmetry energy beyond saturation.

\section*{Acknowledgment}
This work is supported by the National Natural Science
Foundation of China (Grants No. 12305148, No. 12505159, No. 12475149, No. 12175109),
Hebei Natural Science Foundation (No. A2023203055),
Fundamental Research Program of ShanxiProvince (No. 202303021211135),
Natural Science Foundation of Shandong Province under Grants No. ZR2023QA112,
and Fundamental Research Funds for the Central Universities under Grant No. 26CX03025A.


\begin{thebibliography}{ }

\bibitem{Lattimer2007}
J. M. Lattimer and M. Prakash,
\href{https://doi.org/10.1016/j.physrep.2007.02.003}{Phys. Rep. \textbf{442}, 109 (2007)}.
\bibitem{Steiner2005}
A. W. Steiner, M. Prakash, J. M. Lattimer, and P. J. Ellis,
\href{https://doi.org/10.1016/j.physrep.2005.06.002}{Phys. Rep. \textbf{411}, 325 (2005)}.
\bibitem{Lattimer2021}
J. M. Lattimer,
\href{https://doi.org/10.1146/annurev-nucl-102419-124827}{Annu. Rev. Nucl. Part. Sci. \textbf{71}, 433 (2021)}.
\bibitem{Lattimer2001}
J. M. Lattimer and M. Prakash,
\href{https://doi.org/10.1086/319702}{Astrophys. J. \textbf{550}, 426 (2001)}.

\bibitem{Arzoumanian2018}
Z. Arzoumanian, A. Brazier, S. Burke-Spolaor, S. Chamberlin, S. Chatterjee, B. Christy, J. M. Cordes, N. J. Cornish, F. Crawford, and H. T. Cromartie,
\href{https://doi.org/10.3847/1538-4365/aab5b0}{Astrophys. J. Suppl. Ser. \textbf{235}, 37 (2018)}.
\bibitem{Antoniadis2013}
J. Antoniadis, P. C. C. Freire, N. Wex, T. M. Tauris, R. S. Lynch, M. H. van Kerkwijk, M. Kramer, C. Bassa, V. S. Dhillon, T. Driebe, et al.,
\href{https://doi.org/10.1126/science.1233232}{Science \textbf{340}, 1233232 (2013)}.
\bibitem{Riley2021}
T. E. Riley, A. L. Watts, P. S. Ray, S. Bogdanov, S. Guillot, S. M. Morsink, A. V. Bilous, Z. Arzoumanian, D. Choudhury, J. S. Deneva, et al.,
\href{https://doi.org/10.3847/2041-8213/ac0a81}{Astrophys. J. Lett. \textbf{918}, L27 (2021)}.
\bibitem{Miller2019}
M. C. Miller, F. K. Lamb, A. J. Dittmann, S. Bogdanov, Z. Arzoumanian, K. C. Gendreau, S. Guillot, A. K. Harding, W. C. G. Ho, and J. M. Lattimer,
\href{https://doi.org/10.3847/2041-8213/ab50c5}{Astrophys. J. Lett. \textbf{887}, L24 (2019)}.
\bibitem{Riley2019}
T. E. Riley, A. L. Watts, S. Bogdanov, P. S. Ray, R. M. Ludlam, S. Guillot, Z. Arzoumanian, C. L. Baker, A. V. Bilous, D. Chakrabarty, et al.,
\href{https://doi.org/10.3847/2041-8213/ab481c}{Astrophys. J. Lett. \textbf{887}, L21 (2019)}.
\bibitem{Vinciguerra2024}
S. Vinciguerra, T. Salmi, A. L. Watts, D. Choudhury, T. E. Riley, P. S. Ray, S. Bogdanov, Y. Kini, S. Guillot, D. Chakrabarty, et al.,
\href{https://doi.org/10.3847/1538-4357/acfb83}{Astrophys. J. \textbf{961}, 62 (2024)}.
\bibitem{Miller2021}
M. C. Miller, F. K. Lamb, A. J. Dittmann, S. Bogdanov, Z. Arzoumanian, K. C. Gendreau, S. Guillot, W. C. G. Ho, J. M. Lattimer, M. Loewenstein, et al.,
\href{https://doi.org/10.3847/2041-8213/ac089b}{Astrophys. J. Lett. \textbf{918}, L28 (2021)}.
\bibitem{Salmi2024a}
T. Salmi, D. Choudhury, Y. Kini, T. E. Riley, S. Vinciguerra, A. L. Watts, M. T. Wolff, Z. Arzoumanian, S. Bogdanov, D. Chakrabarty, et al.,
\href{https://doi.org/10.3847/1538-4357/ad5f1f}{Astrophys. J. \textbf{974}, 294 (2024)}.
\bibitem{Choudhury2024}
D. Choudhury, T. Salmi, S. Vinciguerra, T. E. Riley, Y. Kini, A. L. Watts, B. Dorsman, S. Bogdanov, S. Guillot, and P. S. Ray,
\href{https://doi.org/10.3847/2041-8213/ad5a6f}{Astrophys. J. Lett. \textbf{971}, L20 (2024)}.
\bibitem{Salmi2024b}
T. Salmi, J. S. Deneva, P. S. Ray, A. L. Watts, D. Choudhury, Y. Kini, S. Vinciguerra, H. T. Cromartie, M. T. Wolff, Z. Arzoumanian, et al.,
\href{https://doi.org/10.3847/1538-4357/ad81d2}{Astrophys. J. \textbf{976}, 58 (2024)}.
\bibitem{Mauviard2025}
L. Mauviard, S. Guillot, T. Salmi, D. Choudhury, B. Dorsman, D. Gonz\'{a}lez-Caniulef, M. Hoogkamer, D. Huppenkothen, C. Kazantsev, and Y. Kini,
\href{https://doi.org/10.3847/1538-4357/ae145d}{Astrophys. J. \textbf{995}, 60 (2025)}.
\bibitem{Miller2026}
M. C. Miller, A. J. Dittmann, I. M. Holt, F. K. Lamb, C. Chirenti, Z. Arzoumanian, J. Berteaud, S. Bogdanov, K. C. Gendreau, A. K. Harding, et al.,
\href{https://doi.org/10.48550/arXiv.2609.00965}{arXiv:2609.00965 (2026)}.

\bibitem{Abbott2018}
B. P. Abbott et al.,
\href{https://doi.org/10.1103/PhysRevLett.121.161101}{Phys. Rev. Lett. \textbf{121}, 161101 (2018)}.
\bibitem{Abbott2019}
B. P. Abbott et al.,
\href{https://doi.org/10.1103/PhysRevX.9.011001}{Phys. Rev. X \textbf{9}, 011001 (2019)}.
\bibitem{Malik2018}
T. Malik, N. Alam, M. Fortin, C. Provid\^{e}ncia, B. K. Agrawal, T. K. Jha, B. Kumar, and S. K. Patra,
\href{https://doi.org/10.1103/PhysRevC.98.035804}{Phys. Rev. C \textbf{98}, 035804 (2018)}.

\bibitem{Margueron2018}
J. Margueron, R. Hoffmann Casali, and F. Gulminelli,
\href{https://doi.org/10.1103/PhysRevC.97.025806}{Phys. Rev. C \textbf{97}, 025806 (2018)}.
\bibitem{Wu2019}
X. Wu and H. Shen,
\href{https://doi.org/10.1103/PhysRevC.99.065802}{Phys. Rev. C \textbf{99}, 065802 (2019)}.
\bibitem{Ji2019}
F. Ji, J. Hu, S. Bao, and H. Shen,
\href{https://doi.org/10.1103/PhysRevC.100.045801}{Phys. Rev. C \textbf{100}, 045801 (2019)}.
\bibitem{Hu2020}
J. Hu, S. Bao, Y. Zhang, and H. Shen,
\href{https://doi.org/10.1093/ptep/ptaa016}{PTEP \textbf{2020}, 043D01 (2020)}.
\bibitem{Wu2021}
X. Wu, S. Bao, H. Shen, and R. Xu,
\href{https://doi.org/10.1103/PhysRevC.104.015802}{Phys. Rev. C \textbf{104}, 015802 (2021)}.
\bibitem{Kubis2023}
S. Kubis, W. W\'ojcik, D. A. Castillo, and N. Zabari,
\href{https://doi.org/10.1103/PhysRevC.108.045803}{Phys. Rev. C \textbf{108}, 045803 (2023)}.
\bibitem{Wu2026}
X. Wu, S. Bao, M. Ju, J. Hu, and H. Shen,
\href{https://doi.org/10.1051/0004-6361/202452766}{A\&A \textbf{706}, A132 (2026)}.

\bibitem{Li2008}
B. A. Li, L. W. Chen, and C. M. Ko,
\href{https://doi.org/10.1016/j.physrep.2008.04.005}{Phys. Rep. \textbf{464}, 113 (2008)}.
\bibitem{Tsang2012}
M. B. Tsang, J. R. Stone, F. Camera, P. Danielewicz, S. Gandolfi, K. Hebeler, C. J. Horowitz, J. Lee, W. G. Lynch, Z. Kohley, et al.,
\href{https://doi.org/10.1103/PhysRevC.86.015803}{Phys. Rev. C \textbf{86}, 015803 (2012)}.
\bibitem{Lattimer2014}
J. M. Lattimer,
\href{https://doi.org/10.1016/j.nuclphysa.2014.04.008}{Nucl. Phys. A \textbf{928}, 276 (2014)}.
\bibitem{Baldo2016}
M. Baldo and G. F. Burgio,
\href{https://doi.org/10.1016/j.ppnp.2016.06.006}{Prog. Part. Nucl. Phys. \textbf{91}, 203 (2016)}.
\bibitem{Yeasin2025}
P. S. Yeasin, S. Tsiopelas, A. Sedrakian, and J.-J. Li,
\href{https://doi.org/10.1103/mz9y-hqlw}{Phys. Rev. D \textbf{112}, 063028 (2025)}.
\bibitem{Bao2014}
S. Bao, J. Hu, Z. Zhang, and H. Shen,
\href{https://doi.org/10.1103/PhysRevC.90.045802}{Phys. Rev. C \textbf{90}, 045802 (2014)}.
\bibitem{Burgio2024}
G. F. Burgio, H. C. Das, and I. Vida\~{n}a,
\href{https://doi.org/10.3389/fspas.2024.1505560}{Front. Astron. Space Sci. \textbf{11}, 1505560 (2024)}.
\bibitem{Ju2025}
M. Ju, X. Wu, and H. Shen,
\href{https://doi.org/10.1103/PhysRevC.111.055801}{Phys. Rev. C \textbf{111}, 055801 (2025)}.
\bibitem{Tews2017}
I. Tews, J. M. Lattimer, A. Ohnishi, and E. E. Kolomeitsev,
\href{https://doi.org/10.3847/1538-4357/aa8db9}{Astrophys. J. \textbf{848}, 105 (2017)}.
\bibitem{Zhang2018}
N. B. Zhang, B. A. Li, and J. Xu,
\href{https://doi.org/10.3847/1538-4357/aac027}{Astrophys. J. \textbf{859}, 90 (2018)}.
\bibitem{Li2019}
B. A. Li, P. G. Krastev, D. H. Wen, and N. B. Zhang,
\href{https://doi.org/10.1140/epja/i2019-12780-8}{Eur. Phys. J. A \textbf{55}, 117 (2019)}.
\bibitem{Krastev2023}
P. G. Krastev,
\href{https://doi.org/10.3390/sym15051123}{Symmetry \textbf{15}, 1123 (2023)}.
\bibitem{Lattimer2023}
J. M. Lattimer,
\href{https://doi.org/10.3390/particles6010003}{Particles \textbf{6}, 30 (2023)}.
\bibitem{Holt2018}
J. W. Holt and Y. Lim,
\href{https://doi.org/10.1016/j.physletb.2018.07.038}{Phys. Lett. B \textbf{784}, 77 (2018)}.
\bibitem{An2023}
R. An, S. Sun, L. G. Cao, and F. S. Zhang,
\href{https://doi.org/10.1007/s41365-023-01269-1}{Nucl. Sci. Tech. \textbf{34}, 119 (2023)}.
\bibitem{Koliogiannis2026}
P. S. Koliogiannis, E. Y\"uksel, T. Ghosh, and N. Paar,
\href{https://doi.org/10.3847/2041-8213/ae261d}{Astrophys. J. Lett. \textbf{996}, L18 (2026)}.
\bibitem{Fattoyev2012}
F. J. Fattoyev, W. G. Newton, J. Xu, and B. A. Li,
\href{https://doi.org/10.1103/PhysRevC.86.025804}{Phys. Rev. C \textbf{86}, 025804 (2012)}.
\bibitem{Kruger2013}
T. Kr\"{u}ger, I. Tews, K. Hebeler, and A. Schwenk,
\href{https://doi.org/10.1103/PhysRevC.88.025802}{Phys. Rev. C \textbf{88}, 025802 (2013)}.
\bibitem{Dong2025}
X. Dong, H. Shen, J. Hu, and Y. Zhang,
\href{https://doi.org/10.1103/hlgb-47pj}{Phys. Rev. D \textbf{112}, 063043 (2025)}.
\bibitem{Li2026}
B. A. Li, X. Grundler, W. J. Xie, and N. B. Zhang,
\href{https://doi.org/10.3847/1538-4357/ae38c0}{Astrophys. J. \textbf{998}, 262 (2026)}.

\bibitem{Li2020}
B. A. Li and M. Magno,
\href{https://doi.org/10.1103/PhysRevC.102.045807}{Phys. Rev. C \textbf{102}, 045807 (2020)}.
\bibitem{Li2021}
B. A. Li, B. J. Cai, W. J. Xie, and N. B. Zhang,
\href{https://doi.org/10.3390/universe7060182}{Universe \textbf{7}, 182 (2021)}.


\bibitem{sugahara1994relativistic}
Y. Sugahara and H. Toki,
\href{https://doi.org/10.1016/0375-9474(94)90923-7}{Nucl. Phys. A \textbf{579}, 557 (1994)}.
\bibitem{fattoyev2010relativistic}
F. J. Fattoyev, C. J. Horowitz, J. Piekarewicz, and G. Shen,
\href{https://doi.org/10.1103/PhysRevC.82.055803}{Phys. Rev. C \textbf{82}, 055803 (2010)}.

\bibitem{Mondal2017}
C. Mondal, B. K. Agrawal, J. N. De, S. K. Samaddar, M. Centelles, and X. Vi\~{n}as,
\href{https://doi.org/10.1103/PhysRevC.96.021302}{Phys. Rev. C \textbf{96}, 021302(R) (2017)}.











%
%
%
\end{thebibliography}

\end{document}